\documentclass{PoS}

\usepackage[T1]{fontenc} 
\usepackage[utf8]{inputenc}
\usepackage{slashed}
\usepackage{amssymb}
\usepackage{amsmath}

\newcommand{\beq}{\begin{eqnarray}}
\newcommand{\eeq}{\end{eqnarray}}

\newcommand{\as}{\alpha_{\mbox{\tiny{S}}}}
\newcommand{\eps}{\epsilon}

\newcommand{\sig}{\sigma}

\newcommand{\gam}{\gamma}

\newcommand{\npo}{{n+1}}
\newcommand{\npt}{{n+2}}

\newcommand{\mc}{\mathcal}

\newcommand{\one}{\, (\mathbf{1})}
\newcommand{\two}{\, (\mathbf{2})}

\newcommand{\RV}{(\mathbf{RV})}
\newcommand{\LO}{{\mbox{\tiny{LO}}}}
\newcommand{\NLO}{{\mbox{\tiny{NLO}}}}
\newcommand{\NNLO}{{\mbox{\tiny{NNLO}}}}

\newcommand{\nnb}{\nonumber}

\newcommand{\kt}{\tilde{k}}

\def\eq#1{Eq.~(\ref{#1})}

\newcommand{\bS}[1]{{\bf S}_{#1}}
\newcommand{\bC}[1]{{\bf C}_{#1}}
\newcommand{\bSC}[1]{{\bf SC}_{#1}}

\newcommand{\bbL}[2]{\overline{\bf L}^{#1}_{#2}}
\newcommand{\bbS}[1]{\overline{\bf S}_{#1}}
\newcommand{\bbC}[1]{\overline{\bf C}_{#1}}
\newcommand{\bbSC}[1]{\overline{\bf SC}_{#1}}

\newcommand{\W}[1]{\mc{W}_{#1}}

\newcommand{\kkl}[1]{\{\bar k\}^{(#1)}}
\newcommand{\kk}[2]{\bar k_{#1}^{(#2)}}
\newcommand{\sk}[2]{\bar s_{#1}^{(#2)}}

\newcommand{\Norm}{\mc{N}_1}

\newcommand{\Bn}{B}

\newcommand{\RR}{RR}

\newcommand{\I}{I}
\newcommand{\bra}{\langle}
\newcommand{\ket}{\rangle}

\title{Local analytic sector subtraction for final state radiation at NNLO}

\ShortTitle{Local analytic sector subtraction for final state radiation at NNLO}

\author{Lorenzo Magnea,
            Ezio Maina,
            Giovanni Pelliccioli,
            Chiara Signorile-Signorile,
            Paolo Torrielli and
            Sandro Uccirati\footnote{Speaker}\\
      Department of Physics, University of Torino, and INFN, Sezione di Torino\\
      E-mail: \email{uccirati@to.infn.it}}

\abstract{
We present recent developments of the {\em local analytic sector} 
subtraction of infrared singularities for final state real radiation at 
NNLO in QCD.
}

\FullConference{
14th International Symposium on Radiative Corrections (RADCOR2019)
\\ 
9-13 September 2019 
\\
Palais des Papes, Avignon, France
}

\begin{document}


\section{Introduction}
\label{intro}

\noindent
The Large Hadron Collider (LHC) is entering its high-precision phase, and 
theoretical predictions need to achieve similar degree of accuracy, in 
order to have the Standard Model background under control and be able to 
disentangle possible signals of new physics. 
As the LHC is a hadron machine, basically all processes are essentially 
QCD-based, and precise theoretical predictions must take into account 
higher-order effects in QCD perturbation theory. 
To this end, many ingredients are necessary: 
an accurate determination of parton distribution functions, a proper 
description of final-state hadronic jets, as well as resummations to 
all orders of large fixed-order contributions. 
But of course the main ingredient to have accurate predictions is the 
computation of QCD corrections to the relevant partonic processes, at a 
sufficiently high order, which actually means at next-to-next-to-leading 
order (NNLO) for the most common processes. 
Furthermore, because of the variety and complexity of scattering processes, 
it would be desirable that these NNLO QCD computations could be automated, 
at the same level presently achieved at next-to-leading order (NLO). 
To reach this ambitious goal, one needs not only automated computations of 
two-loop corrections, but also a universal framework to deal with the 
cancellation of soft and collinear singularities, arising both in virtual 
corrections and in the phase space integration of unresolved real 
radiation of massless particles. 

\noindent
The most precise way to treat these cancellation is by means of a 
{\it subtraction procedure}, which basically consists in subtracting from 
the real squared matrix elements one or more simple local counterterms, 
mimicking its singular behaviour in the entire phase space, and adding 
them back, integrated in the extra radiations, in order to cancel the 
singularities of the virtual matrix element. 
There is a lot of freedom in defining these counterterms and in the way 
the integration of the radiated phase space is performed, giving raise to 
many possible subtraction procedures. 
At NLO, the most successful general algorithms are the 
{\it Frixione-Kunzst-Signer (FKS)}~\cite{Frixione:1995ms}, the 
{\it Catani-Seymour (CS)}~\cite{Catani:1996vz} and the 
{\it Nagy-Soper}~\cite{Nagy:2003qn} subtraction methods. 
At NNLO, the overlapping of singular regions increases the complexity of 
the problem, and several different methods, not always based on a 
subtraction procedure, have been developed, however, so far, without 
reaching the desired degree of generality and automation. 
The first subtraction procedure to be developed at NNLO was the {\it Antenna} 
subtraction~\cite{GehrmannDeRidder:2005cm}, which is essentially a 
generalisation of the NLO CS subtraction. 
A different framework, based on the known singular limits of the squared 
matrix elements with double real radiation, is the {\it CoLoRFulNNLO} 
subtraction~\cite{Somogyi:2005xz}. 
A complete numerical approach, extending the FKS subtraction at NNLO, is 
the {\it Sector-improved residue} subtraction~\cite{Czakon:2010td}, which 
basically generalises the subtraction procedures based on the sector 
decomposition technique~\cite{Binoth:2000ps,Anastasiou:2003gr}, and was 
the first method to be successfully applied to a hadronic scattering 
process (top pair production) at NNLO.
A recent analytical development of this approach is the 
{\it Nested Soft-Collinear} subtraction~\cite{Caola:2017dug}.
Successful methods not based on a local subtraction procedure are the 
$q_T$ slicing~\cite{Catani:2007vq} and {\it N-Jettiness} 
slicing~\cite{Boughezal:2015dva}.
Finally, new methods, or refinements of existing ones, are also being 
introduced~\cite{Cacciari:2015jma,Sborlini:2016hat,Herzog:2018ily}.

\noindent
Recently, we presented a new approach~\cite{Magnea:2018hab}, which we 
called {\it local analytic sector subtraction}: it attempts to take 
maximal advantage of the available freedom in the definition of the local 
infrared counterterms, essentially combining ideas that have been 
successfully implemented at NLO.
The first crucial element is the partition of phase space in sectors, 
as done in the {\it FKS} subtraction~\cite{Frixione:1995ms}, by means of 
sector functions obeying  a set of sum rules which allow to simplify the 
analytic integration of counterterms when sectors are appropriately 
recombined. 
A second key ingredient is the remapping of momenta to Born kinematics, 
following CS factorisation of the radiative phase space, which is 
particularly suitable for a straightforward integration of counterterms 
at NLO. 
Finally, we use the known expressions for the 2-unresolved singular 
limits~\cite{Catani:1999ss} and take maximal advantage of the simple 
structure of multiple limits, which follows from the factorisation 
properties of scattering amplitudes~\cite{Magnea:2018ebr}. 


\section{Local analytic sector subtraction for final state radiation at NLO}

\noindent
At NLO the differential cross sections $d\sig_\NLO/dX$ with respect to any 
infrared-safe observable $X$ can schematically be written as
\beq
\label{eq:NLOstruct}
\frac{d \sig_\NLO - d \sig_\LO}{d X} 
& = & 
\int d\Phi_n \, V \, \delta_n (X) 
+ 
\int d\Phi_\npo \, R \, \delta_\npo (X) 
\, .
\eeq
where $R$ and $V$ denote the real and virtual squared matrix elements 
respectively, the latter renormalised in the $\overline{\rm MS}$ scheme. 
We have also introduced $\delta_i(X)\equiv\delta(X-X_i)$, with $X_i$ 
representing the observable $X$ computed with $i$-body kinematics.
In dimensional regularisation, with $d = 4 - 2 \eps$ space-time dimensions, 
the virtual contribution features poles in $\eps$, while the real 
contribution is characterised by singularities in the radiation phase 
space, which are of soft and collinear nature. 
When computed in $d$ dimensions, the phase space integration in $d\Phi_\npo$ 
results in explicit $\eps$ poles, which cancel those of virtual 
origin~\cite{Kinoshita:1962ur,Lee:1964is}. 

\noindent
Any local subtraction procedure at NLO consists in adding and subtracting 
a counterterm $K$ to \eq{eq:NLOstruct}, and exploiting the factorisation 
of the $(n+1)$ phase space $d\Phi_\npo = d\Phi_n\,d\Phi_1$, getting 
\beq
\label{eq:subNLO}
\frac{d\sig_\NLO-d\sig_\LO}{dX} 
& = & 
\! \int \! d \Phi_n \Big[ V +  I \Big] \, 
\delta_n (X)
+ 
\! \int \! d\Phi_\npo 
\Big[ R \, \delta_\npo (X) - K \, \delta_n (X) \Big]
\, ,
\quad \, 
I =
\! \int \! d\Phi_1  \, K 
\, .
\qquad
\eeq
The counterterm $K$ must reproduce all the singular limits of the 
real-radiation contribution $R$, so that the combination $R-K$ does not 
present any phase space singularities. 
Its integral $I$ in the radiative phase space $d\Phi_1$ features poles in 
$\eps$, which exactly cancel those of the virtual squared matrix element 
$V$. 
The choice of the counterterm $K$ and of the phase space factorisation 
$d\Phi_\npo = d\Phi_n\,d\Phi_1$ defines the subtraction scheme. 

\noindent
In our local analytic sector subtraction scheme for final state radiation, 
we first introduce the FKS sector functions $\W{ij}$, forcing the 
projection $R\,\W{ij}$ to approach a singular configuration only if the 
final-state particle $i$ becomes soft, or particles $i$ and $j$ become 
collinear. 
Requiring for the sector functions the sum rule 
\beq
\label{eq: sumrule}
\sum_{i,j\neq i} \W{ij} & = & 1
\, ,
\eeq
we can construct the counterterm $K$ as 
\beq
\label{eq: K}
K
& = &
\sum_{i, \, j \neq i} 
\Big[ 
\Big( \bbS{i} \, R \, \W{ij} \Big)
+
\Big( \bbC{ij} \, R \, \W{ij} \Big)
- 
\Big( \bbS{i}\bbC{ij} \, R \,\W{ij} \Big)
\Big]
\, ,
\eeq
where we have introduced the operators $\bbS{i}$ and $\bbC{ij}$, which act 
on all objects to their right in the following way: 
$\bbS{i}$ and $\bbC{ij}$ extract the leading behaviour for particle $i$ 
becoming soft and for particles $i$ and $j$ becoming collinear, 
respectively; 
when acting on matrix elements, they also define implicitly a remapping of 
momenta (to be specified), such that the resolved particles of all matrix 
elements are on the mass-shell and satisfy four-momentum conservation. 
Concretely  
\beq
\label{eq: S,C,SC}
\bbS{i} \,R
& = &
-\,\Norm \,
\sum_{\substack{c\neq i\\d\neq i,c}}
\frac{s_{cd}}{s_{ic}s_{id}} \, 
B_{cd}\!\left(\kkl{icd}\right)
\, ,
\qquad\qquad
\bbS{i} \bbC{ij} \,R
\, = \,
2\,\Norm \, C_{f_j} \, 
\frac{s_{jr}}{s_{ij}s_{ir}} \, 
B\!\left(\kkl{ijr}\right)
\, ,
\nnb
\eeq
\vspace{-5mm}
\beq
\bbC{ij} \, R 
& = & 
\frac{\Norm}{s_{ij}} \, \left[
P_{ij} \,
{\Bn} \Big( \! \kkl{ijr} \! \Big)
+
Q_{ij}^{\mu\nu} \,
{\Bn}_{\mu\nu} \Big( \! \kkl{ijr} \! \Big)
\right]
\, ,
\eeq
where $s_{ab}=2k_a\cdot k_b$, 
$\Norm = 8\pi\as(\mu^2 e^{\gam_E}/(4\pi))^{\eps}$, $B_{cd}$ is the 
colour-connected Born-level squared matrix element, and $B_{\mu\nu}$ is 
the spin-connected Born-level squared matrix element.
The spin-averaged Altarelli-Parisi kernels $P_{ij}$ and the azimuthal 
kernels $Q_{ij}^{\mu\nu}$ are functions of $x_i=s_{ir}/(s_{ir}+s_{jr})$ 
and $x_j=s_{jr}/(s_{ir}+s_{jr})$, defined by 
\beq
P_{ij} 
& = & 
\delta_{f_i g} \delta_{f_j g} \, 2 \, C_A 
\left( \frac{x_i}{x_j} + \frac{x_j}{x_i} + x_i x_j \right) 
+ 
\delta_{ \{f_i f_j\} \{q \bar q\} } \, T_R
\left( 1 - \frac{2 x_i x_j}{1 - \eps} \right) 
\nnb \\
&& 
+ \, 
\delta_{f_i \{q, \bar q\}} \delta_{f_j g} \, C_F 
\left( \frac{1 + x_i^2}{x_j} - \eps x_j \right) 
+ 
\delta_{f_i g} \delta_{f_j \{q, \bar q\}} \, C_F
\left( \frac{1 + x_j^2}{x_i} - \eps x_i \right) 
\, ,
\nnb \\
Q_{ij}^{\mu\nu} 
& = & 
\left[
- \, 
\delta_{f_ig} \, \delta_{f_jg} \, 2 \, C_A \,  x_i x_j 
+ 
\delta_{\{f_i f_j\}\{q \bar q\}} \, T_R \, \frac{2 x_i x_j}{1 - \eps} 
\right]
\left[ - \, g^{\mu\nu} + (d - 2) \, \frac{\kt_i^\mu \kt_i^\nu}{\kt_i^2}   \right] 
\, .
\eeq

\noindent
The next important step is the choice of the remappings $\kkl{ijr}$ and 
$\kkl{icd}$ that, in our approach, are not referred to the specific 
sector, as in FKS, but depend on the IR kernels of \eq{eq: S,C,SC}. 
We decided to use CS remappings, which are particularly suited for an 
easy analytic integration of the counterterms, defined by 
\beq
\label{eq: CS1}
\kk{b}{abc} \, = \, k_{a} + k_{b} - \frac{s_{ab}}{s_{ac} + s_{bc}} \, k_{c}
\, ,
\qquad
\kk{c}{abc} \, = \, \frac{s_{abc}}{s_{ac} + s_{bc}} \, k_{c}
 \, ,
\qquad
\kk{i}{abc} \, = \, k_i,
\quad
\mbox{if } i \neq a,b,c
\, ,
\qquad
\eeq
where $s_{abc} = s_{ab} + s_{ac} + s_{bc}$. 
Under these remappings, the $(n+1)$-particle phase space factorises as 
\vspace{-4mm}
\beq 
d\Phi_\npo 
& = & 
d\Phi_n\Big( \! \kkl{abc} \! \Big) \,
d\Phi_1\Big( \sk{bc}{abc}; y, z, \phi \Big)
\nnb
\eeq
\vspace{-5mm}
\beq
\int d \Phi_1 \left(s; y, z, \phi \right)
& \equiv & 
N_1 \, s^{1 - \eps} \!
\int_0^\pi \!\! d \phi \, \sin^{- 2 \eps} \! \phi \int_0^1 \!\! dy 
\int_0^1 \!\! dz
\Big[ y (1 - y)^2 \, z (1 - z) \Big]^{- \eps} \! (1 - y) \, ,
\nnb
\eeq
where the invariants are given by 
\beq
s_{ab} \, = \, y \, \sk{bc}{abc}, 
\quad
s_{ac} \, = \, z (1 - y) \, \sk{bc}{abc}, 
\quad
s_{bc} \, = \, (1 - z)(1 - y) \, \sk{bc}{abc}, 
\quad
\sk{bc}{abc} = 2\kk{b}{abc}\!\cdot\kk{c}{abc} \, ,
\quad
\eeq
and $N_1=(4\pi)^{\eps - 2}\pi^{-1/2}/\Gamma(1/2-\eps)$. 
The integral $I$ of the counterterm $K$ in the $d\Phi_1$ phase space can 
then be computed analytically, after having summed away the sector 
functions, obtaining 
\beq
I
& = &
\frac{\as}{2 \pi} \,
\bigg[
\;
\sum_{c,d\neq c} \!
J_{\rm s}^{cd} \,
B_{cd}
+
\sum_{p}
J_{\rm hc}^{pr} \,
B
\;
\bigg] 
\, ,
\\[-6mm]
\nnb
\eeq
where
\beq
J_{\rm s}^{cd}
& = &
- \frac{1}{\eps^2} 
- \frac{2}{\eps} 
- 6 
+ \frac{7}{2} \, \zeta_2
+
\ln\frac{\bar s_{cd}}{\mu^2}
\left( \frac{1}{\eps} + 2 - \frac12\,\ln\frac{\bar s_{cd}}{\mu^2} \right)
+
\mc O(\eps)
\, ,
\nnb \\
J_{\rm hc}^{pr}
& = &
- \,
\delta_{f_p g} \, \frac{C_A + 4\,T_R\,N_f}{6}
\left( \frac{1}{\eps} + \frac{8}{3} - \ln\frac{\bar s_{pr}}{\mu^2} \right)
- 
\delta_{f_p \{q, \bar q\}} \, \frac {C_F}{2}
\left( \frac{1}{\eps} + 2 - \ln\frac{\bar s_{pr}}{\mu^2} \right) 
+
\mc O(\eps)
\, .
\quad
\eeq


\section{Local analytic sector subtraction for final state radiation at NNLO}

\noindent
At NNLO the structure of the differential cross section contains three 
contributions, 
\beq
\label{eq:NNLOstruct}
\frac{d \sig_\NNLO - d \sig_\NLO}{d X} 
& = & 
\int d\Phi_n \, VV \, \delta_n (X) 
+ 
\int d\Phi_\npo \, RV \, \delta_\npo (X) 
+ 
\int d\Phi_\npt \, RR \, \delta_\npt (X) 
\, .
\eeq
where $RR$, $RV$ and $VV$ denote the double-real, real-virtual and 
double-virtual squared matrix elements respectively, the latter two 
renormalised in the $\overline{\rm MS}$ scheme. 
The double-virtual contribution has only poles in $\eps$, while the 
real-virtual contribution features both poles in $\eps$ and phase space 
singularities, and the double-real term is characterised only by 
singularities in the radiation phase space. 
When computed in $d$ dimensions, the phase space integrations in 
$d\Phi_\npo$ and $d\Phi_\npt$ result in explicit poles in $\eps$, which 
cancel those arising from virtual 
corrections~\cite{Kinoshita:1962ur,Lee:1964is}. 

\noindent
In this case the structure of the counterterms is more involved than at 
NLO and we construct it step by step. 
Following the same strategy described in the previous section, we first 
introduce new sector functions $\W{ijkl}$ for $RR$, while we use  the NLO 
sector functions $\W{ij}$ for $RV$. 
The singular behaviour of $RV$ for soft and/or collinear emission is 
similar to $R$ at NLO, and we build the corresponding counterterm in the 
same way, according to 
\beq
\label{eq: K^RV}
K^{\RV}
& = &
\sum_{i, \, j \neq i} 
\Big[ 
\Big( \bbS{i} \, RV \, \W{ij} \Big)
+
\Big( \bbC{ij} \, RV \, \W{ij} \Big)
- 
\Big( \bbS{i}\bbC{ij} \, RV \,\W{ij} \Big)
\Big]
\, ,
\eeq
The soft and/or collinear singular behaviour of $RV$ is 
known~\cite{Somogyi:2006db}, and the explicit expressions for $\bbS{i}RV$, 
$\bbC{ij}RV$, and $\bbS{i}\bbC{ij}RV$ is obtained by introducing proper 
remappings in the matrix elements, analogous to those introduced at NLO. 

\noindent
The new sector functions $\W{ijkl}$ for $RR$ are defined to minimize the 
number of singular regions of $RR\,\W{ijkl}$, and must of course sum to 
$1$, according to 
\beq
\sum_{i,j\neq i} \, \sum_{\substack{k \neq i\\l\neq i,k}}
\W{ijkl} & = & 1
\, .
\eeq
Notice that in the previous formula we allow the last two indices $k$ and 
$l$ to be equal to the second index $j$. 
This is done to catch specific collinear limits of $RR$: 
in $RR\,\W{ijjk}$ and $RR\,\W{ijkj}$ ($k \neq j$), the only singular 
double collinear limit is when the three particles $i,j,k$ become 
simultaneously collinear (which we represent by the operator $\bC{ijk}$); 
on the other hand, in $RR\,\W{ijkl}$ ($k,l \neq j$) the surviving singular 
double collinear limit is when the two pairs  $(i,j)$ and $(k,l)$ become 
collinear separately (which we represent by the operator $\bC{ijkl}$). 
It is possible to define the sector functions $\W{ijkl}$ such that in the 
three mentioned topologies just the following singular limits survive: 
\beq
\W{ijjk}\,RR 
&\; :\quad& 
\bS{i},
\quad
\bC{ij},
\quad
\bS{ij},
\quad
\bC{ijk},
\quad
\bSC{ijk},
\qquad\qquad\quad\;\;\;
j \neq i, \; k \neq i,j;
\nnb \\
\W{ijkj}\,RR
&\; :\quad& 
\bS{i},
\quad
\bC{ij},
\quad
\bS{ik},
\quad
\bC{ijk},
\quad
\bSC{ijk},
\quad
\bSC{kij},
\qquad
j \neq i, \; k \neq i,j;
\nnb \\
\W{ijkl}\,RR
&\; :\quad& 
\bS{i},
\quad
\bC{ij},
\quad
\bS{ik},
\quad
\bC{ijkl},
\quad
\bSC{ikl},
\quad
\bSC{kij},
\qquad
j \neq i, \; k \neq i,j, \; l \neq i,j,k.
\qquad
\eeq
where, besides the operators for the single soft limit $\bS{i}$ and for the 
single collinear limit $\bC{ij}$, and the already mentioned operators for 
the double collinear limits $\bC{ijk}$ and $\bC{ijkl}$, we have defined 
the double-soft operator for particles $i$ and $j$, $\bS{ij}$, and the 
soft-collinear operator $\bSC{ijk}$, which extract the singular 
behaviour when particle $i$ becomes soft and simultaneously particles 
$j$ and $k$ become collinear. 
Of course this list of limits depends on the actual form of the sector 
functions. 

\noindent
Since all previous limits commute when acting on both $RR$ and sector 
functions, we can easily build expressions which, by construction have no 
phase space singularties. 
Indeed 
\beq
\label{eq:finite}
&&
(1-\bbS{i})
(1-\bbC{ij})
(1-\bbS{ij})
(1-\bbC{ijk})
(1-\bbSC{ijk})
\,
\RR \; \W{ijjk}
\; = \;
\mbox{finite}
\, ,
\\
&&
(1-\bbS{i})
(1-\bbC{ij})
(1-\bbS{ik})
(1-\bbC{ijk})
(1-\bbSC{ijk})
(1-\bbSC{kij})
\,
\RR \; \W{ijkj}
\; = \;
\mbox{finite}
\, ,
\nnb \\
&&
(1-\bbS{i})
(1-\bbC{ij})
(1-\bbS{ik})
(1-\bbC{ijkl})
(1-\bbSC{ikl})
(1-\bbSC{kij})
\,
\RR \; \W{ijkl}
\; = \;
\mbox{finite}
\quad 
(k,l \neq j)
\, ,
\nnb
\eeq
where the bar denotes again an implicit remapping on matrix elements, to 
preserve mass-shell conditions and momuntum conservation. 
The explicit expressions for $\bbS{i}RR$ and $\bbC{ij}RR$ are analogous 
to the ones at NLO, while $\bbC{ijkl}RR$ and $\bbSC{ijk}RR$ are 
essentially products of two single operators and can be obtained again 
from the NLO case (see Ref.~\cite{Magnea:2018hab} for the explicit 
expressions). 
The only non-trivial limits are $\bbS{ij}RR$ and $\bbC{ijk}RR$, which 
are given by 
\beq
\label{eq:SS CC}
\bbS{ij} \, \RR 
&=&
\frac{\Norm^{\,2}}{2} \!
\sum_{\substack{c \neq i,j \\ d \neq i,j,c}} \!
\Bigg[
\sum_{\substack{e \neq i,j,c,d \\ f \neq i,j,c,d}} \!\!\!\!\!
\I_{cd}^{(i)} \,
\I_{ef}^{(j)} 
{\Bn}_{cdef} \Big( \! \kkl{icd,jef} \! \Big) 
+
4 \!\!\!\!\!
\sum_{\substack{e \neq i,j,c,d}} \!\!\!\!\!
\I_{cd}^{(i)} \,
\I_{ed}^{(j)} 
{\Bn}_{cded} \Big( \! \kkl{icd,jed} \! \Big) 
\nnb\\
&&
\qquad
+ \,
2\,
\I_{cd}^{(i)} \, 
\I_{cd}^{(j)}
{\Bn}_{cdcd} \Big( \! \kkl{ijcd} \! \Big) 
+
\left( 
\I_{cd}^{(ij)} - \frac12\,\I_{cc}^{(ij)} - \frac12\,\I_{dd}^{(ij)} 
\right) \!
{\Bn}_{cd} \Big( \! \kkl{ijcd} \! \Big) 
\Bigg]
\nnb \\
\bbC{ijk} \, \RR 
& = & 
\frac{\Norm^2}{s_{ijk}^2} \, \left[
P_{ijk} \,
{\Bn} \Big( \! \kkl{ijkr} \! \Big)
+
Q_{ijk}^{\mu\nu} \,
{\Bn}_{\mu\nu} \Big( \! \kkl{ijkr} \! \Big)
\right]
\eeq
where $\I_{cd}^{(i)}= s_{cd}/(s_{ic}s_{id})$ is the NLO eikonal factor, 
while $\I_{cd}^{(ij)}$, $P_{ijk}$, and $Q_{ijk}^{\mu\nu}$ are pure NNLO soft 
and collinear kernels, which have been computed explicitly in 
Ref.~\cite{Catani:1999ss}, and will be analysed in more details in the 
next section. 
The remappings introduced implicitly in \eq{eq:SS CC} are again chosen to 
simplify the analytical integration procedure, and are basically double CS 
remappings, given by 
\beq
\label{eq: CS2}
\kk{c}{abcd} 
\, = \, 
k_{a} + k_{b} + k_{c} - \frac{s_{abc}}{s_{ad} + s_{bd} + s_{cd}} \, k_{d} 
\, ,
\qquad
\kk{d}{abcd} \, = \, \frac{s_{abcd}}{s_{ad} + s_{bd} + s_{cd}} \, k_{d}
 \, ,
\eeq
while $\kk{i}{abcd} \, = \, k_i$ if $i \neq a,b,c,d$. 

\noindent
From the finite expressions of \eq{eq:finite}, we construct the 
counterterms which cancel the phase space singularities of $RR$. 
To this end we introduce the 1- and 2-unresolved limits $\bbL{\one}{ij}$ 
and $\bbL{\two}{ijkl}$, as 
\beq
1-\bbL{\one}{ij} 
& \equiv &
(1-\bbS{i})
(1-\bbC{ij})
\, ,
\nnb \\
1-\bbL{\two}{ijjk} 
& \equiv &
(1-\bbS{ij})
(1-\bbC{ijk})
(1-\bbSC{ijk})
\nnb \\
1-\bbL{\two}{ijkj} 
& \equiv &
(1-\bbS{ik})
(1-\bbC{ijk})
(1-\bbSC{ijk})
(1-\bbSC{kij})
\, ,
\nnb \\
1-\bbL{\two}{ijkl} 
& \equiv &
(1-\bbS{ik})
(1-\bbC{ijkl})
(1-\bbSC{ikl})
(1-\bbSC{kij})
\, ,
\qquad
\mbox{for}\quad k,l \neq j
\, .
\eeq
In this way we can rewrite the three equations (\ref{eq:finite}) in one 
formula (for $k \neq i,l\neq i,k$)
\beq
( 1-\bbL{\one}{ij} )
( 1-\bbL{\two}{ijkl} )
\,
RR \; \W{ijkl}
\; = \;
\Big[ 
RR 
-
\bbL{\one}{ij}
- 
\bbL{\two}{ijkl}
+
\bbL{\one}{ij}\,\bbL{\two}{ijkl}
\Big]
\; 
\W{ijkl}
\; = \;
\mbox{finite}
\, ,
\eeq
and define the three counterterms 
\beq
K^{\bf (1)}
& = &
\sum_{i,j\neq i} \, \sum_{\substack{k \neq i\\l\neq i,k}}
\bbL{\one}{ij} \, RR \, \W{ijkl}
\, ,
\qquad\qquad
K^{\bf (2)} 
\; = \;
\sum_{i,j\neq i} \, \sum_{\substack{k \neq i\\l\neq i,k}}
\bbL{\two}{ijkl} \, RR \, \W{ijkl}
\, ,
\nnb
\eeq
\vspace{-9mm}
\beq
K^{\bf (12)} 
& = &
-
\sum_{i,j\neq i} \, \sum_{\substack{k \neq i\\l\neq i,k}}
\bbL{\one}{ij} \, \bbL{\two}{ijkl} \, RR \, \W{ijkl}
\, .
\eeq
Finally, we can build our subtraction formula, which we write as 
\beq
\label{eq:subNNLO}
\frac{d\sig_\NNLO-d\sigma_\NLO}{dX}
&=&
\int\! d\Phi_n \left( VV + I^{\two} + I^{\RV} \right)\,\delta_{n}(X)
\nnb \\
&&
+
\int\! d\Phi_\npo \left[
\left( RV + I^{\one} \right) \delta_{\npo}(X)
-
\left( K^{\RV} - I^{\bf (12)} \right)\delta_{n}(X)
\right]
\nnb \\
&&
+
\int\! d\Phi_\npt \left[
RR\,\delta_{\npt}(X)
-
K^{\one}\delta_{\npo}(X)
-
\left( K^{\two} + K^{\bf (12)} \right)\delta_{n}(X)
\right]
\, .
\qquad
\eeq
where $I^{\one}$, $I^{\two}$, $I^{\bf (12)}$, and $I^{\RV}$ are given by 
\beq
I^{\one}
\!\! & = & \!\!\!
\int d\Phi_1 \, K^{\one},
\quad
I^{\bf (12)}
= \!
\int d\Phi_1 \, K^{\bf (12)},
\quad
I^{\two}
= \!
\int d\Phi_2 \, K^{\two},
\quad
I^{\RV}
= \!
\int d\Phi_1 \, K^{\RV}.
\qquad
\eeq
The calculation of $I^{\one}$ and $I^{\bf (12)}$ needs the integrations in 
a single radiation phase space $d\Phi_1$, and can be readily performed 
following the NLO case, obtaining 
\beq
I^{\one} 
& = & 
\frac{\as}{2 \pi} \,
\sum_{k,l\neq k} \!
\bigg[
\sum_{c,d\neq c} \!
J_{\rm s}^{cd} \,
R_{cd}
+
\sum_{p}
J_{\rm hc}^{pr} \,
R
\,
\bigg]
\mc W_{kl}
\, ,
\nnb \\
I^{\bf (12)} 
& = & 
-\,
\frac{\as}{2 \pi} \,
\sum_{k,l\neq k} \!
\Big[ \bbS{k} + \bbC{kl}(1-\bbS{k}) \Big]
\bigg[
\sum_{c,d\neq c} \!
J_{\rm s}^{cd} \,
R_{cd}
+
\sum_{p}
J_{\rm hc}^{pr} \,
R
\,
\bigg]
\mc W_{kl}
\, .
\eeq
As one can see, $I^{\bf (12)}$ corresponds to the IR limit of 
$I^{\bf (1)}$, with opposite sign. 
The second line of \eq{eq:subNNLO} is therefore free from phase space 
singularities, exactly as the third one. 
Explicit calculations show that $I^{\one}$ cancels the $\eps$ 
poles of $RV$ and $I^{\bf (12)}$ cancels those of $K^{\RV}$.

\noindent
Of course, because of the KLN theorem~\cite{Kinoshita:1962ur,Lee:1964is}, 
$I^{\two}$ and $I^{\RV}$ cancel the $\eps$ poles of $VV$. 
Their integration is the most difficult part of the calculation, but can 
be performed following the procedure sketched in the next session. 

\subsection{Integration of $I^{\two}$ and $I^{\RV}$}

\noindent
The integrals $I^{\two}$ and $I^{\RV}$ consist of many terms. 
A large fraction of these  terms are convolutions of integrals of the NLO 
type and their integration is trivial, but we refrain from showing them 
here for the sake of brevity. 
We describe instead the method we used to integrate the most intricate 
parts  of these counterterms, namely those which depend also on the 
azimuthal angle of the unresolved particle(s). 
To make the integration of such terms feasible, the choice of remappings 
and the factorisation of phase space are crucial. 
Single and double CS remappings seem the best choice in view of the 
analytical integration, because they involve only the invariants that are 
actually present in the singular kernels and moreover generate a simple 
radiative phase space. 

\noindent
For $I^{\RV}$, involving at most 5 invariants $s_{ab}$, $s_{ac}$, $s_{bc}$, 
$s_{cd}$, $s_{ad}$ (where $a$ is the unresolved particle), the CS 
remappings of \eq{eq: CS1} give: 
\beq 
d\Phi_\npo 
& = & 
d\Phi_n\Big( \! \kkl{abc} \! \Big) \,
d\Phi_1\Big( \sk{bc}{abc}; y, z, x \Big)
\, ,
\nnb
\eeq
\vspace{-6mm}
\beq
\int d \Phi_1\Big(s; y, z, x \Big)
\!\!
&=&
\!\!
2^{-2\eps}\,N_1\,s^{1-\eps}\!
\int_0^1\!\!dx \!\int_0^1\!\!\!dy \!\int_0^1\!\!\!dz\,
[x(1\!-\!x)]^{-\eps-\frac12}\,
\Big[ y (1 - y)^2 z (1 - z) \Big]^{- \eps} \! (1 - y) 
\, ,
\nnb
\eeq
\vspace{-6mm}
\beq
&&
s_{ab} 
\, = \,
y\,\sk{bc}{abc}
\, ,
\quad
s_{ac}
\, = \,
z\,(1-y)\,\sk{bc}{abc}
\, ,
\quad
s_{bc}
\, = \,
(1-z)(1-y)\,\sk{bc}{abc}
\, ,
\quad
s_{dc}
\, = \,
(1-y)\,\sk{cd}{abc}
\, ,
\nnb\\[+2mm]
&&
s_{ad}
\, = \,
y\,(1-z) \,\sk{cd}{abc}
+ 
z \,\sk{bd}{abc}
- 
2\,(1-2x)\sqrt{y\,z\,(1-z)\,\sk{bd}{abc}\,\sk{cd}{abc}}
\, .
\qquad
\eeq
For $I^{\two}$, involving at most 6 invariants $s_{ab}$, $s_{ac}$, $s_{bc}$, 
$s_{cd}$, $s_{ad}$, $s_{bd}$ (where $a$ and $b$ are the unresolved 
particles), the double CS remappings of \eq{eq: CS2} give  
\beq 
d\Phi_\npt 
& = & 
d\Phi_n\Big( \! \kkl{abcd} \! \Big) \,
d\Phi_2\Big( \sk{cd}{abcd}; y', z', x', y, z, \phi \Big)
\, ,
\nnb
\eeq
\vspace{-6mm}
\beq
\int \! d\Phi_2( s; y', z', x', y, z, \phi)
& = & 
N_2 \, s^{2 - 2 \eps} \!\!
\int_0^1 \!\!\!\! d x' \!
\int_0^1 \!\!\!\! d y' \!
\int_0^1 \!\!\!\! d z' \!
\int_0^\pi \!\!\!\! d \phi (\sin\phi)^{\!- 2 \eps} \!
\int_0^1 \!\!\!\! d y \!
\int_0^1 \!\!\!\! d z \, 
\big[ x'(1-x')\big]^{-\eps-\frac12} 
\nnb\\
& & 
\times\, 
\Big[ y'(1-y')^2\,z'(1-z')\,y^2(1-y)^2\,z(1-z) \Big]^{- \eps} 
(1-y') \, y(1-y) 
\, ,
\nnb
\eeq
\vspace{-6mm}
\beq
s_{ab} 
& = & 
y' \, y \, \sk{cd}{abcd} 
\, ,
\qquad
s_{ac} 
\, = \, 
z' ( 1 - y' ) \, y \, \sk{cd}{abcd} 
\, ,
\qquad
s_{bc} 
\, = \, 
( 1 - y' ) ( 1  - z' ) \, y \, \sk{cd}{abcd} 
\, ,
\nnb \\[+2mm]
s_{bd} 
& = & 
(1-y) \left[\,
y'z'(1-z) + (1-z')z + 2(1-2x')\sqrt{y'z'(1-z')z(1-z)} 
\,\right] \sk{cd}{abcd} 
\, ,
\nnb \\
s_{cd} 
& = &
( 1 - y' ) ( 1 - y ) ( 1 - z ) \, \sk{cd}{abcd} 
\, ,
\qquad
s_{ad} 
\, = \,
( y' + z - y' z) ( 1 - y ) \, \sk{cd}{abcd} 
-
s_{bd}
\, ,
\eeq
with $N_2=2^{- 2 \eps}N_1^2$. 

\noindent
In order to explain how the integration is performed, we restrict the 
analysis to the azimuth-dependent terms of $I^{\two}$, namely to the 
integration  in $d\Phi_2$ of $\I_{cd}^{(ij)}$ and $P_{ijk}$ of 
\eq{eq:SS CC} (but note that with the techniques shown here we were 
able to integrate $I^{\two}$ and $I^{\RV}$ completely). 

\noindent
The explicit expression of $\I_{cd}^{(ij)}$ is 
\beq
\I_{cd}^{(ij)}
&=&
\delta_{\{f_if_j\}\{q \bar q\}}\,\frac{T_R}{2}\,
\I_{ijcd}^{(q\bar q)}
-
\delta_{f_i g}\,\delta_{f_j g}\frac{C_A}{2}\,
\I_{ijcd}^{(gg)}
\eeq
where 
$\delta_{\{f_af_b\}\{q\bar q\}}=\delta_{f_a q}\delta_{f_b\bar q}+\delta_{f_a\bar q}\delta_{f_b q}$,  
while $\I_{ijcd}^{(q\bar q)}$ and $\I_{ijcd}^{(gg)}$ are taken from 
Ref.~\cite{Catani:1999ss}, 
\beq
\begin{array}{llll}
\I_{ijcd}^{(q\bar q)}
& = &
\I_{cd}(k_i,k_j)
& \quad
\mbox{(eq. (96) of~\cite{Catani:1999ss})}
\, ,
\nnb\\[2mm]
\I_{ijcd}^{(gg)}
& = &
S_{cd}(k_i,k_j)
& \quad
\mbox{(eq. (110) of~\cite{Catani:1999ss})}
\, .
\end{array}
\eeq
On the other hand, for $P_{ijk}$ we have 
\beq
P_{ijk} 
&=& 
  P_{ijk}^{(qq'\bar q')}\,\delta_{\{f_if_j\}\{q \bar q\}}\,\delta_{f_k \{q'\bar q'\}}
+ P_{jki}^{(qq'\bar q')}\,\delta_{\{f_jf_k\}\{q \bar q\}}\,\delta_{f_i \{q'\bar q'\}}
+ P_{kij}^{(qq'\bar q')}\,\delta_{\{f_kf_i\}\{q \bar q\}}\,\delta_{f_j \{q'\bar q'\}}
\nnb\\
&&
+ \; P_{ijk}^{(qq\bar q)}\,\delta_{\{f_i\{f_jf_k\}\}\{q \bar q\}}
+    P_{jki}^{(qq\bar q)}\,\delta_{\{f_j\{f_kf_i\}\}\{q \bar q\}}
+    P_{kij}^{(qq\bar q)}\,\delta_{\{f_k\{f_if_j\}\}\{q \bar q\}}
\nnb\\
&&
+ \; P_{ijk}^{(gq\bar q)}\,\delta_{\{f_if_j\}\{q \bar q\}}\,\delta_{f_k g}
+    P_{jki}^{(gq\bar q)}\,\delta_{\{f_jf_k\}\{q \bar q\}}\,\delta_{f_i g}
+    P_{kij}^{(gq\bar q)}\,\delta_{\{f_kf_i\}\{q \bar q\}}\,\delta_{f_j g}
\nnb\\
&&
+ \; P_{ijk}^{(ggq)}\,\delta_{f_i g}\,\delta_{f_j g}\,\delta_{f_k \{q,\bar q\}}
+    P_{jki}^{(ggq)}\,\delta_{f_j g}\,\delta_{f_k g}\,\delta_{f_i \{q,\bar q\}}
+    P_{kij}^{(ggq)}\,\delta_{f_k g}\,\delta_{f_i g}\,\delta_{f_j \{q,\bar q\}}
\nnb\\
&&
+ \, P_{ijk}^{(ggg)}\,\delta_{f_i g}\,\delta_{f_j g}\,\delta_{f_k g} \, ,
\eeq
where $q$ and $q'$ are quarks of equal or different flavours and 
\beq
\delta_{f_a\{q\bar q\}} = \delta_{f_a q} + \delta_{f_a \bar q}
\, , 
\qquad
\delta_{\{f_a\{f_bf_c\}\}\{q \bar q\}} = 
\delta_{f_a q}\,\delta_{f_b \bar q}\,\delta_{f_c \bar q}
+ 
\delta_{f_a \bar q}\,\delta_{f_b q}\,\delta_{f_c q}
\, .
\eeq
The expressions for $P_{ijk}^{(qq'\bar q')}$, $P_{ijk}^{(qq\bar q)}$, 
$P_{ijk}^{(gq\bar q)}$, $P_{ijk}^{(ggq)}$ and $P_{ijk}^{(ggg)}$ can again 
be found in~\cite{Catani:1999ss}:
\beq
\begin{array}{llll}
P_{ijk}^{(qq'\bar q')} 
& = &
\bra \hat P_{\bar q_i q_j q_k'} \ket 
&
\mbox{(eq. (57) of~\cite{Catani:1999ss})}
\, ,
\nnb\\[2mm]
P_{ijk}^{(qq\bar q)} 
& = &
\bra \hat P_{\bar q_i q_j q_k}^{(\rm id)} \ket 
&
\mbox{(eq. (59) of~\cite{Catani:1999ss})}
\, ,
\nnb\\[2mm]
P_{ijk}^{(gq\bar q)} 
& = &
C_F \, T_R\,
\bra \hat P_{g_k q_i \bar q_j}^{(\rm ab)} \ket 
+
C_A \, T_R\,
\bra \hat P_{g_k q_i \bar q_j}^{(\rm nab)} \ket 
&
\mbox{(eqs. (68) and (69) of~\cite{Catani:1999ss})}
\, ,
\nnb\\[2mm]
P_{ijk}^{(ggq)} 
& = &
C_F^2 \,
\bra \hat P_{g_i g_j q_k}^{(\rm ab)} \ket 
+
C_F \, C_A \,
\bra \hat P_{g_i g_j q_k}^{(\rm nab)} \ket 
&
\mbox{(eqs. (61) and (62) of~\cite{Catani:1999ss})}
\, ,
\nnb\\[2mm]
P_{ijk}^{(ggg)} 
& = &
\bra \hat P_{g_i g_j g_k} \ket 
&
\mbox{(eq. (70) of~\cite{Catani:1999ss})}
\, .
\end{array}
\eeq
From the expressions of $\I_{cd}^{(ij)}$ and $P_{ijk}$ we see that they 
are symmetric under the permutation of some of the involved momenta. 
However, when integrating in the two-body radiative phase space 
$d\Phi_2$, we have a larger freedom of choosing the outgoing momenta 
$k_a$, $k_b$, $k_c$, $k_d$, depending on the symmetries of their four-body 
phase space, which is invariant under 
\begin{itemize}
\item
any permutations of the four momenta $k_a$, $k_b$, $k_c$, $k_d$;
\item
any of the following permutations of invariants: 
$s_{ab} \leftrightarrow s_{cd}$, $s_{ac} \leftrightarrow s_{bd}$,
and $s_{ad} \leftrightarrow s_{bc}$.
\end{itemize}
These symmetries reflect in the factorisation of phase space: in fact, when 
reparametrising the four body phase space from $(k_a, k_b, k_c, k_d)$ to 
$(\kk{c}{abcd}, \kk{d}{abcd}, y,z,\phi,y',z',x')$, we have the freedom 
of performing any one of the permutations listed above.
This is of crucial importance in simplifying the analytical computation of 
the $d\Phi_2$ integration of $\I_{cd}^{(ij)}$ and $P_{ijk}$. 
Exploiting this freedom in a systematic way, it is possible to rearrange 
$\I_{cd}^{(ij)}$ and $P_{ijk}$ so that in the denominators of each 
term only the following combinations of invariants appear: 
\[
s_{ab}, \quad s_{ac}, \quad s_{bc}, \quad s_{bd}, \quad s_{cd}, 
\quad 
s_{ac}+s_{bc}, \quad s_{ad}+s_{bd}, \quad s_{ab}+s_{bc}.
\]
Among these denominators, only $s_{bd}$ depends on the azimuthal angle 
(parametrised by the variable $x'$): 
therefore all terms without $s_{bd}$ in the denominator can be trivially 
integrated in $dx'$, and those with $1/s_{bd}$ can be integrated using the 
integral relation 
\beq
I_{b}(A,B)
&\equiv& 
\int_0^1 \!\!\! d x' \, 
\big[ x'(1-x')\big]^{\frac12-b}
\frac{1}{A^2+B^2+2(1-2x')AB}
\\
&=& 
\frac{\Gamma^2\!\left(\frac32\!-\!b\right)}{\Gamma(3\!-\!2b)}\,
\bigg[
\frac{1}{B^2}\,{}_2F_1\!\bigg(1,b,2-b,\frac{A^2}{B^2}\bigg)
\Theta(B^2\!-\!A^2)
+ \,
\frac{1}{A^2}\,{}_2F_1\!\bigg(1,b,2-b,\frac{B^2}{A^2}\bigg)
\Theta(A^2\!-\!B^2)
\bigg]
\, .
\nnb
\eeq
In addition to the integration in $x'$, we could perform those in $\phi$ 
(there is no $\phi$ dependence in the integrands) and $y$ (giving just 
Beta functions). 
The $z$ and $z'$ integrations are performed by using known properties of 
the hypergeometric function ${}_2F_1$, and by introducing integrals in a 
new variable $t$ (with no direct physical meaning). 
The remaining integrations are then of the following types: 
\beq
\!\!\!\!
&&
\int_0^1 \!\!\! dt
\!
\int_0^1 \!\!\! dy'
\,
(1-t)^{\mu} \, t^{\nu}\,
(1-y')^{\rho} \, (y')^{\sigma}
{}_2F_1(n_1,n_2-\eps,n_3-2\eps,1-ty')
\, ,
\\
\!\!\!\!
&&
\int_0^1 \!\!\! dt
\,
(1-t)^{\mu} \, t^{\nu}\,
{}_2F_1(n_1,n_2-\eps,n_3-2\eps,1-t)
\, ,
\qquad\quad
n_1,n_2,n_3 \in\mathbb{N}, 
\quad
n_1 \ge 1, 
\quad
n_3 \ge n_1+1,n_2
\, ,
\nnb
\eeq
with $\mu,\nu,\rho,\sigma = n + m\,\eps$ ($n,m\in\mathbb{Z}$, $n \ge -1$). 
These integrals can of course be written in terms of hypergemetric 
functions ${}_2F_1$, ${}_3F_2$, ${}_4F_3$. 
Since however we are not interested in the full $\eps$ dependence, we 
first expanded in $\eps$ and then integrated in $t$ and $y'$, 
obtaining the following compact results 
{\small
\beq
\int\!\!d\Phi_2( s; y', z', x', y, z, \phi) \, \I_{ijcd}^{(X)} 
\, = \,
A(s) \, \mc I_{ijcd}^{(X)}
\, ,
\qquad\quad
\int\!\!d\Phi_2( s; y', z', x', y, z, \phi) \, P_{ijk}^{(X)} 
\, = \,
A(s) \, \mc P_{ijk}^{(X)}
\, ,
\qquad
\eeq
}
with $A(s) = s^{-2\eps}e^{-2\eps\gamma_{_E}}/(4\pi)^{4-2\eps}$ and 
{\scriptsize
\beq
\mc I_{ijcd}^{(q\bar q)}
\!\! & = & \!\!
\frac{2}{3}  \, \frac1{\eps^3}
+ 
\frac{28}{9} \, \frac1{\eps^2}
+ 
\left[ \frac{416}{27} - \frac{7}{9} \pi^2 \right] \! \frac1{\eps}
+ 
\frac{5260}{81} 
- 
\frac{104}{27} \pi^2 
- 
\frac{76}{9} \zeta(3)
\qquad
(c \neq d)
\, ,
\\
\mc I_{ijcc}^{(q\bar q)} 
\!\! & = & \!\!
- \frac{2}{3} \, \frac1{\eps^2}
- \frac{16}{9} \, \frac1{\eps}
- \frac{212}{27} + \pi^2
\, ,
\nnb\\
\mc I_{ijcd}^{(gg)}
\!\! & = & \!\!
\frac2{\eps^4}
+ 
\frac{35}{3} \, \frac1{\eps^3}
+
\left[ \frac{481}{9} - \frac{8}{3} \pi^2 \right] \! \frac1{\eps^2}
+
\left[ \frac{6218}{27} - \frac{269}{18} \pi^2 - \frac{154}{3} \zeta(3) \right] \!
\frac1{\eps}
+
\frac{76912}{81} 
-
\frac{3775}{54} \pi^2 
-
\frac{2050}{9} \zeta(3) 
-
\frac{23}{60} \pi^4
\qquad
(c \neq d)
\, ,
\nnb\\
\mc I_{ijcc}^{(gg)}
\!\! & = & \!\!
-
\frac{2}{3}  \, \frac1{\eps^2}
-
\frac{10}{9} \, \frac1{\eps}
-
\frac{164}{27} 
+ 
\pi^2
\, ,
\nnb
\eeq
\vspace{-5mm}
\beq
\mc P_{ijk}^{(qq'\bar q')}
\!\!\! & = & \!\!
-
\frac{1}{3}  \, \frac1{\eps^3}
-
\frac{31}{18} \, \frac1{\eps^2}
+
\left[ - \frac{889}{108} + \frac{\pi^2}{2} \right] \! \frac1{\eps}
-
\frac{23941}{648} 
+
\frac{31}{12} \pi^2 
+ 
\frac{80}{9} \zeta(3)
\, ,
\\
\mc P_{ijk}^{(qq\bar q)}
\!\!\! & = & \!\!
\left[ - \frac{13}{8} + \frac{1}{4} \pi^2 - \zeta(3) \right] \!
\frac1{\eps}
-
\frac{227}{16}
+
\pi^2 
+ 
\frac{17}{2} \zeta(3)
-
\frac{11}{120} \pi^4
\, ,
\nnb\\
\mc P_{ijk}^{(gq\bar q)}
\!\!\! & = & \!\!
C_A T_R \,
\Bigg\{
-
\frac{2}{3}  \, \frac1{\eps^3}
- 
\frac{41}{12} \, \frac1{\eps^2}
+
\left[ - \frac{1675}{108} + \frac{17}{18} \pi^2 \right] \! \frac1{\eps}
-
\frac{5404}{81} 
+
\frac{1063}{216} \pi^2 
+ 
\frac{139}{9} \zeta(3)
\Bigg\}
\nnb\\
&&
+ \,
C_F T_R \,
\Bigg\{
-
\frac{2}{3}  \, \frac1{\eps^3}
-
\frac{31}{9} \, \frac1{\eps^2}
+
\left[ - \frac{889}{54} + \pi^2 \right] \! \frac1{\eps}
-
\frac{23833}{324} 
+
\frac{31}{6} \pi^2 
+ 
\frac{160}{9} \zeta(3)
\Bigg\}
\, ,
\nnb\\
\mc P_{ijk}^{(ggq)}
\!\!\! & = & \!\!
C_F C_A \,
\Bigg\{
\frac{1}{2} \, \frac1{\eps^4}
+
\frac{8}{3} \, \frac1{\eps^3}
+
\left[ \frac{905}{72} - \frac{2}{3} \pi^2 \right] \! \frac1{\eps^2}
+
\left[ 
\frac{11773}{216} - \frac{89}{24} \pi^2 - \frac{65}{6} \zeta(3) 
\right] \!
\frac1{\eps}
+ 
\frac{295789}{1296} 
-
\frac{845}{48} \pi^2 
-
\frac{2191}{36} \zeta(3)
+
\frac{19}{240} \pi^4
\!
\Bigg\}
\nnb\\
&&
+ \,
C_F^2 \,
\Bigg\{
\frac2{\eps^4}
+
\frac7{\eps^3}
+
\left[ \frac{251}{8} - 3 \pi^2 \right] \! \frac1{\eps^2}
+
\left[ 
\frac{2125}{16} - \frac{21}{2} \pi^2 - \frac{154}{3} \zeta(3) 
\right] \! \frac1{\eps}
+
\frac{17607}{32} 
-
\frac{753}{16} \pi^2 
-
\frac{548}{3} \zeta(3)
+
\frac{13}{20} \pi^4 
\Bigg\}
\, ,
\nnb\\
\mc P_{ijk}^{(ggg)}
\!\!\! & = & \!\!
\frac{5}{2} \, \frac1{\eps^4}
+
\frac{21}{2} \, \frac1{\eps^3}
+
\left[ \frac{853}{18} - \frac{11}{3} \pi^2 \right] \! \frac1{\eps^2}
+
\left[ 
\frac{5450}{27} - \frac{275}{18} \pi^2 - \frac{188}{3} \zeta(3) 
\right] \!
\frac1{\eps}
+
\frac{180739}{216} 
-
\frac{1868}{27} \pi^2 
-
\frac{1555}{6} \zeta(3)
+
\frac{41}{60} \pi^4
\, .
\nnb
\eeq
}
As a cross check, all these integrals have been computed also numerically, 
using sector decomposition and without using the symmetries of the phase 
space. 


\section{Summary}

\noindent
We have presented the latest developments of the 
{\em local analytic sector} subtraction. 
The method takes advantage of the partition of phase space through 
sector functions. 
This, in turn, allows to easily identify counterterms by using the known 
singular limits of the real matrix elements, and by introducing proper 
remappings of the momenta, in order to preserve mass-shell conditions 
and momentum conservation at each step of the calculation. 
We exploit the freedom in the choice of remappings, and obtain a simple 
factorisation of phase space, that gives us the possibility to integrate 
analytically all counterterms in the radiation phase space. 
We have shown the application of the method to final state radiation, 
sketching the procedure developed for integrating the counterterms, and 
giving analytic results for the integrated double-soft and 
double-collinear kernels. 




\begin{thebibliography}{99}

\bibitem{Frixione:1995ms}
S.~Frixione, Z.~Kunszt, and A.~Signer,
  {\it {Three jet cross-sections to next-to-leading order}},
  {\em Nucl. Phys.} {\bf B467} (1996) 399--442,
  [\href{http://xxx.lanl.gov/abs/hep-ph/9512328}{{\tt hep-ph/9512328}}].

\bibitem{Catani:1996vz}
S.~Catani and M.~H. Seymour,
  {\it {A General algorithm for calculating jet cross-sections in NLO QCD}},
  {\em Nucl. Phys.} {\bf B485} (1997) 291--419,
  [\href{http://xxx.lanl.gov/abs/hep-ph/9605323}{{\tt hep-ph/9605323}}].
  [Erratum: Nucl. Phys.B510,503(1998)].

\bibitem{Nagy:2003qn} 
  Z.~Nagy and D.~E.~Soper, 
  {\it {General subtraction method for numerical calculation of one loop QCD matrix elements}}, 
  {\em JHEP} {\bf 0309}, 055 (2003)
  [\href{http://xxx.lanl.gov/abs/hep-ph/0308127}{{\tt hep-ph/0308127}}].

\bibitem{GehrmannDeRidder:2005cm} 
  A.~Gehrmann-De Ridder, T.~Gehrmann and E.~W.~N.~Glover,
  {\it {Antenna subtraction at NNLO}},
  {\em JHEP} {\bf 0509}, 056 (2005),
  [\href{http://xxx.lanl.gov/abs/hep-ph/0505111}{{\tt hep-ph/0505111}}].

\bibitem{Somogyi:2005xz} 
  G.~Somogyi, Z.~Trocsanyi and V.~Del Duca,
  {\it {Matching of singly- and doubly-unresolved limits of tree-level QCD squared matrix elements}},
  {\em JHEP} {\bf 0506}, 024 (2005),
  [\href{http://xxx.lanl.gov/abs/hep-ph/0502226}{{\tt hep-ph/0502226}}].

\bibitem{Czakon:2010td}
  M.~Czakon,
  {\it {A novel subtraction scheme for double-real radiation at NNLO}},
  {\em Phys. Lett.} {\bf B693} (2010) 259--268,
  [\href{http://xxx.lanl.gov/abs/1005.0274}{{\tt 1005.0274}}].

\bibitem{Binoth:2000ps} 
  T.~Binoth and G.~Heinrich,
  {\it {An automatized algorithm to compute infrared divergent multiloop integrals}},
  {\em Nucl.\ Phys.\ B} {\bf 585}, 741 (2000),
  [\href{http://xxx.lanl.gov/abs/hep-ph/0004013}{{\tt hep-ph/0004013}}].

\bibitem{Anastasiou:2003gr} 
  C.~Anastasiou, K.~Melnikov and F.~Petriello,
  {\it {A new method for real radiation at NNLO}},
  {\em Phys.\ Rev.\ D} {\bf 69}, 076010 (2004),
  [\href{http://xxx.lanl.gov/abs/hep-ph/0311311}{{\tt hep-ph/0311311}}].

\bibitem{Caola:2017dug}
  F.~Caola, K.~Melnikov, and R.~R{\"o}ntsch,
  {\it {Nested soft-collinear subtractions in NNLO QCD computations}},
  {\em Eur. Phys. J.} {\bf C77}  (2017), no.~4 248,
  [\href{http://xxx.lanl.gov/abs/1702.01352}{{\tt 1702.01352}}].

\bibitem{Catani:2007vq}
  S.~Catani and M.~Grazzini,
  {\it {An NNLO subtraction formalism in hadron  collisions and its application to Higgs boson production at the LHC}},
  {\em Phys. Rev. Lett.} {\bf 98} (2007) 222002,
  [\href{http://xxx.lanl.gov/abs/hep-ph/0703012}{{\tt hep-ph/0703012}}].

\bibitem{Boughezal:2015dva} 
  R.~Boughezal, C.~Focke, X.~Liu and F.~Petriello,
  {\it {$W$-boson production in association with a jet at next-to-next-to-leading order in perturbative QCD}},
  {\em Phys.\ Rev.\ Lett.} {\bf 115}, no. 6, 062002 (2015),
  [\href{http://xxx.lanl.gov/abs/1504.02131}{{\tt 1504.02131}}].

\bibitem{Cacciari:2015jma}
  M.~Cacciari, F.~A. Dreyer, A.~Karlberg, G.~P. Salam, and G.~Zanderighi,
  {\it {Fully Differential Vector-Boson-Fusion Higgs Production at Next-to-Next-to-Leading Order}},
  {\em Phys. Rev. Lett.} {\bf 115} (2015),
  no.~8 082002, [\href{http://xxx.lanl.gov/abs/1506.02660}{{\tt 1506.02660}}].

\bibitem{Sborlini:2016hat}
G.~F.~R. Sborlini, F.~Driencourt-Mangin, and G.~Rodrigo,
  {\it {Four-dimensional unsubtraction with massive particles}},
  {\em JHEP} {\bf 10} (2016) 162,
  [\href{http://xxx.lanl.gov/abs/1608.01584}{{\tt 1608.01584}}].

\bibitem{Herzog:2018ily}
  F.~Herzog,
  {\it {Geometric IR subtraction for real radiation}},
  {\em JHEP} {\bf 1808}, 006 (2018),
  \href{http://xxx.lanl.gov/abs/1804.07949}{{\tt 1804.07949}}.

\bibitem{Magnea:2018hab} 
  L.~Magnea, E.~Maina, G.~Pelliccioli, C.~Signorile-Signorile, P.~Torrielli and 
  S.~Uccirati,
  {\it {Local analytic sector subtraction at NNLO}},
  {\em JHEP} {\bf 1812}, 107 (2018), Erratum: [JHEP {\bf 1906}, 013 (2019)]
  \href{http://xxx.lanl.gov/abs/1806.09570}{{\tt 1806.09570}}.

\bibitem{Catani:1999ss} 
  S.~Catani and M.~Grazzini,
  {\it {Infrared factorization of tree level QCD amplitudes at the next-to-next-to-leading order and beyond}},
  {\em Nucl.\ Phys.\ B} {\bf 570}, 287 (2000),
  [\href{http://xxx.lanl.gov/abs/hep-ph/9908523}{{\tt hep-ph/9908523}}].

\bibitem{Magnea:2018ebr} 
  L.~Magnea, E.~Maina, G.~Pelliccioli, C.~Signorile-Signorile, P.~Torrielli and 
  S.~Uccirati,
  {\it {Factorisation and Subtraction beyond NLO}},
  {\em JHEP} {\bf 1812}, 062 (2018)
  \href{http://xxx.lanl.gov/abs/1809.05444}{{\tt 1809.05444}}.

\bibitem{Kinoshita:1962ur}
  T.~Kinoshita,
  {\it {Mass singularities of Feynman amplitudes}},
  {\em J. Math. Phys.} {\bf 3} (1962) 650--677.

\bibitem{Lee:1964is}
  T.~D. Lee and M.~Nauenberg,
  {\it {Degenerate Systems and Mass Singularities}},
  {\em Phys. Rev.} {\bf 133} (1964) B1549--B1562.

\bibitem{Somogyi:2006db} 
  G.~Somogyi and Z.~Trocsanyi,
  {\it {A Subtraction scheme for computing QCD jet cross sections at NNLO: Regularization of real-virtual emission}},
  {\em JHEP} {\bf 0701}, 052 (2007),
  [\href{http://xxx.lanl.gov/abs/hep-ph/0609043}{{\tt hep-ph/0609043}}].

\end{thebibliography}
\end{document}